\documentclass[12pt,preprint]{aastex}
\usepackage{emulateapj5,epsf}
\shorttitle{IMBHs in Spirals}
\shortauthors{Kawakatu et al.}

\begin{document}
%% LaTeX will automatically break titles if they run longer than
%% one line. However, you may use \\ to force a line break if
%% you desire.

\title{Growth of Intermediate Massive Black Holes in the Hierarchical Formation of Small Spiral Galaxies in the High-$z$ Universe}

%% Use \author, \affil, and the \and command to format
%% author and affiliation information.
%% Note that \email has replaced the old \authoremail command
%% from AASTeX v4.0. You can use \email to mark an email address
%% anywhere in the paper, not just in the front matter.
%% As in the title, you can use \\ to force line breaks.

\author{NOZOMU KAWAKATU}
\affil{International School for Advanced Studies (SISSA/ISAS), Via Beirut 2-4, 34014 Trieste, Italy\\ kawakatu@sissa.it}

\author{TAKAYUKI R. SAITOH}
\affil{National Astronomical Observatory of Japan, Mitaka, Tokyo 181-8588, Japan\\ saitoh.takayuki@nao.ac.jp}
  
\and 

\author{KEIICHI WADA\altaffilmark{1}}
\affil{National Astronomical Observatory of Japan, Mitaka, Tokyo 181-8588, Japan\\ wada.keiichi@nao.ac.jp}

%% Notice that each of these authors has alternate affiliations, which
%% are identified by the \altaffilmark after each name.  Specify alternate
%% affiliation information with \altaffiltext, with one command per each
%% affiliation.

\altaffiltext{1}{Department of Astronomical Science, The Graduate University for Advanced Studies, Osawa 2-21-1, Mitaka, Tokyo 181-8588, Japan}

%% Mark off your abstract in the ``abstract'' environment. In the manuscript
%% style, abstract will output a Received/Accepted line after the
%% title and affiliation information. No date will appear since the author
%% does not have this information. The dates will be filled in by the
%% editorial office after submission.

%%%%%%%%%%%%%%%%%%%%%%%%%%%%%%%%%%%
% (2)Abstract  & Subject Headings %
%%%%%%%%%%%%%%%%%%%%%%%%%%%%%%%%%%%

\begin{abstract}
Combining a theoretical model of mass accretion onto a galactic center
with a high-resolution $N$-body/SPH simulation, we investigate
the formation of an intermediate massive black hole (IMBH) during the
hierarchical formation of a small spiral galaxy (with a total mass of
$10^{10}M_{\odot}$) in the high-$z$ universe. We found that the rate of 
average mass accretion to the nucleus due to the radiation drag
exerted by newly formed stars in the forming galaxy is $\approx
10^{-5}M_{\odot}$yr$^{-1}$.
%which is comparable to the Eddington mass@accretion rate for a BH mass wi%th $10^{4}M_{\odot}$. 
As a result of
this accretion, an IMBH with $\approx 10^{4}M_{\odot}$ can be formed
in the center of the spiral galaxy at $z\sim 4$. We found that a central
BH coevolves with the dark matter halo from $z\sim 15$ to $z\sim 2$. The
mass ratio of the BH to the dark matter halo is nearly constant
$\approx (1-3) \times 10^{-6}$ from $z\sim 10$ to $z\sim 2$.  
This is because
that change in the dark matter potential enhances star formation in
the central part of the galaxy, and as a result the BH evolves due to
mass accretion via the radiation drag. Therefore, our model naturally 
predicts a correlation between massive BHs and dark matter halos. 
Moreover, it is found that the final BH-to-bulge mass ratio ($\approx 5\times 10^{-5}$) in a small spiral galaxy at high-$z$ is much smaller than that in the large galaxies ($\approx 10^{-3}$).
Our results also suggest that the scatter in the observed scaling relations between the bulge mass and black hole mass are caused by a time 
lag between BH growth and growth of bulge.  
We also predict that the X-ray luminosity of AGN is
positively correlated with the CO luminosity in the central region.
%For the meanwhile, the BH growing phases clear away from
%the linear relation. Therefore, this features would be one of the
%powerful tools to expose the early stage of BH growth at high-$z$
%universe we have missed so far.
By comparing our results with the properties of Lyman break galaxies (LBGs), it is predicted that some LBGs have massive BHs of 
$\approx 10^{6}-10^{7}M_{\odot}$.

\end{abstract}

%\newpage
%%%%%%%%%%%%%%%%%%%%%%%%%%%%%
% (3)TEXT & Acknowledgments %
%%%%%%%%%%%%%%%%%%%%%%%%%%%%%
\section{Introduction}
%\label{INTRO}
Recent compilation of the kinematical data on galactic centers has revealed that a central 
``massive dark object" (MDO), which is a candidate for a supermassive
black hole (BH), tightly correlates with the mass of a galactic
bulge; the BH-to-bulge mass ratio is $\approx 0.001$ as a median value
(e.g., Kormendy \& Richstone 1995;
Magorrian et al. 1998; Merritt \& Ferrarese 2001; McLure \& Dunlop 2002;
Marconi \& Hunt 2003). 
There have been a number of theoretical efforts to clarify the origin of this
relation (e.g., Silk \& Rees 1998: Ostriker 2000; Adams, Graff \& Richstone 2001).
However, little has been elucidated regarding the physics on the angular momentum transfer in a spheroidal system (a bulge), which is inevitable
for formation of BHs. 
Recently, Ferrarese (2002) and Baes et al. (2003) have argued that the BH mass 
in spiral galaxies is related to the dark matter halo mass.
This correlation suggests that formation of supermassive BHs is physically 
connected not only with formation of galactic bulges, but also 
with assembly processes of dark matter halos in galaxy formation.
Since merging of proto-galaxies triggers active star formation, 
a physical link between star formation and mass accretion toward the central BH is expected.

Umemura (2001) has considered the effects of radiation drag as
a mechanism for removing the angular momentum of the gas in the active galactic nuclei.  
The radiation drag in the solar system is known as the Poynting-Robertson
effect.
Note that, in the early universe, Compton drag force has a similar effect
on formation of massive BHs (Umemura, Loeb, \& Turner 1997). 
The rate of angular momentum loss due to radiation drag is given by
$d \ln J/dt \simeq -\chi_{\rm d} E/c$, 
where $J$ is the total angular momentum of the gaseous component, $E$ is 
the energy density of the uniform spheroidal system, and $\chi_{\rm d}$
is the mass extinction coefficient which is given by $\chi_{\rm
d}=n_{\rm d}\sigma_{\rm d}/\rho_{\rm gas}$ with the number density
$n_{\rm d}$, cross-section $\sigma_{\rm d}$ and gas density $\rho_{\rm
gas}$.
The exact expressions for the radiation drag are given in the literature
(e.g., Umemura, Fukue, \& Mineshige 1997; Fukue, Umemura, \& Mineshige 1997).
 
In an optically thin regime, $d \ln J/dt \simeq -(\tau L_{*}/c^{2}M_{\rm gas})$, 
where $\tau$ is the total optical depth of the system, 
$L_{*}$ is the total luminosity of the spheroidal system, and 
$M_{\rm gas}$ is the total mass of gas. In an optically thick regime, 
the radiation drag efficiency is saturated due to conservation of the photon
number (Tsuribe \& Umemura 1997).
Thus, an expression of the angular momentum loss rate 
suitable for both regimes can be
$d \ln J/dt \simeq -(L_{*}/c^{2}M_{\rm g})(1-e^{-\tau})$.
The mass accretion rate is therefore
$\dot{M}=-M_{\rm g}d \ln J/dt = (L_{*}/c^{2})(1-e^{-\tau})$. 
In an optically thick regime, this gives simply 
$\dot{M}= L_{*}/c^{2}$ (Umemura 2001). 
Thus, the total mass accreted onto the MDO, $M_{\rm MDO}$, is maximally 
$M_{\rm MDO} \simeq \int {L_{*}}/c^{2}dt$.
For more realistic cases, we should take into account inhomogeneity of 
the interstellar matter (ISM). 
In active star-forming galaxies, the ISM is  
observed to be highly inhomogeneous 
(Sanders et al. 1988; Gordon, Calzetti \& Witt 1997). 
In addition, high resolution three-dimensional hydrodynamic simulations have shown
that multiple supernovae (SNe) in a galactic center form a quasi-stable inhomogeneous torus
around a supermassive black hole (Wada \& Norman 2002). 
In such inhomogeneous ISM, optically thin surface layers of optically thick 
clumpy clouds lose their angular momentum due to radiation drag, and
eventually
they accrete toward the galactic center (Sato et al. 2004).
Kawakatu \& Umemura (2002) have shown that the inhomogeneity of 
ISM plays an important role 
in the radiation drag attaining maximal efficiency.
Based on the radiation drag model, 
Kawakatu, Umemura, \& Mori (2003) predict that a mass ratio between 
the black hole mass and the bulge mass, $M_{\rm BH}/M_{\rm bulge}\simeq 0.001$,  
which is determined by the energy conversion efficiency 
of nuclear fusion from hydrogen to helium, i.e., 0.007 (Umemura 2001).
%Then, the final mass of MDO is proportional to 
%the total radiation energy from stars.
%Taking the realistic chemical evolution into consideration, 
%the radiation drag model predicts a mass ratio as
%$M_{\rm BH}/M_{\rm bulge}\simeq 0.001$ in a galactic bulge
%(Kawakatu, Umemura, \& Mori 2003).
%The theoretical upper limit of BH-to-bulge mass ratio 
%is determined by the energy conversion efficiency 
%of nuclear fusion from hydrogen to helium, i.e., 0.007 (Umemura 2001).
In these previous studies, galaxies are treated as a one-zone model, therefore
the growth of supermassive BHs has not been revealed in a more
realistic situation, namely
that of the hierarchical formation of galaxies. 
Granato et al. (2004) presented 
a semi-analytic modeling of the early evolution of 
massive spheroidal galaxies and AGNs within the dark matter halo, 
including the angular momentum transfer via radiation drag. They claimed
the feedback from supernovae and from AGNs determines the relation
between the BH mass, the bulge mass and the dark matter halo mass (see
also Bukert \& Silk 2001).

%On the other hand, direct numerical simulation of galaxy formation can
%be a complementary approach.
Di Matteo et al. (2003) followed the
evolution of the gas, stars, and the dark matter in forming galaxies,
and they found that the observed BH mass-to-stellar velocity dispersion
of a bulge is reproduced if the gas mass in the bulge is linearly
proportional to the black hole mass. However, it is impossible to
resolve the structure of the central sub-kpc region or the bulge
component of host galaxies because of the limitation on their numerical
resolution (mass resolutions are $10^{7}-10^{8}M_{\odot}$ and 
gravitational softening lengths is 4-9 kpc). 
Recently, using high-resolution cosmological $N$-body/SPH simulations with
$2\times 10^{6}$particles (one SPH particle has $10^{3}M_{\odot}$ 
and a gravitational softening length of $\sim$ 50 pc), Saitoh \& Wada
(2004) investigated stellar and gaseous cores on a sub-kpc scale
during the hierarchical formation of a small spiral galaxy (with a total mass
of $10^{10}M_{\odot}$), and found that the galactic core ($<$ several
100 pc) coevolves with the galactic dark matter halo of $\sim$10 kpc scale.
%In this simulation, it has not been clear that the mass accretion from
%central sub-kpc scale to the BH horizon scale because of the limit of
%the spatial resolution, but
The rapid increase of the gas mass and star
formation rate (SFR) in the central sub-kpc region may cause further mass
accretion to the nucleus, due to, for example, a turbulent viscosity
(Wada, Meurer, \& Norman 2002), a gas drag and dynamical friction in
dense stellar clusters (Norman \& Scoville 1988), the radiation drag
originating in the nuclear starburst (e.g., Umemura 2001), or the
BHs-BHs merger (e.g., Haehnelt 2004). 

Here we focus on the radiation drag as one of the possible processes of 
mass accretion onto a BH during hierarchical formation of a galaxy.
%It is suggested that the radiation drag can explain
The observed AGN-starburst connection in nearby galaxies (Heckman et al. 1989; Kauffman et al. 2003; Imanishi \& Wada 2004; Jahnke et al. 2004)
suggests that star formation plays an important role in the mass accretion.
We expect a correlation between BHs and bulges as a natural
result of galaxy formation, if the radiation drag works.  
In this paper, we quantitatively estimate evolution of a black hole mass 
in a forming galaxy, combining $N$-body/SPH simulations 
of cosmological galaxy formation done by Saitoh \& Wada (2004) 
with an analytic model of angular momentum transfer due to the radiation
drag. 

This paper is organized as follows:
In Section 2, we briefly describe the simulation of galaxy formation. 
Our model for the growth of a massive BH via the radiation-hydrodynamic process
is also explained.
Based on this model, in Section 3, we show the history of accretion to a BH in a
spiral galaxy. We also discuss the mutual relationships between a massive BH, 
a galactic bulge, and a dark matter halo. Finally, we discuss
correlation between AGN activities and the properties of bulges. In
Section 4, we compare our predictions with observational scaling relations, and
we discuss the Lyman break galaxies, as candidates for the small spiral
galaxies that we demonstrate here. Section 5 is devoted to conclusions.

\section{Models}
%In this section, we construct the coevolution model by coupling 
%the hierarchical galaxy formation with the mass accretion 
%due to the radiation drag from several 100pc to the galactic center. 
%First, we show a belief introduction of the methods and models for 
%the high-resolution $N$-body/SPH simulations.
%More details of the simulations are given in the literature
%(Saitoh \& Wada 2004; Saitoh et al. in preparation).
%After that, we build up the BH growth based on the radiation drag-driven mass accretion.

\subsection{Simulations of galaxy formation}
The numerical simulations used here are based on Saitoh \& Wada (2004) and
Saitoh et al. (in preparation).
We model the formation and evolution of galaxies in the CDM universe,
adopting a top-hat initial condition with an open boundary 
for a single galactic halo ($M_{\rm halo} \sim 10^{10}M_{\odot}$).
The cosmological parameters in our model are $\Omega_{0}=1.0$, $\Omega_{\lambda}=0.0$, 
$\Omega_{\rm b}=0.1$, $h=H_{0}/{\rm km}/{\rm s}/{\rm Mpc}=0.5$, and $\sigma_{8}=0.63$. 
The collapse epoch of the halo is set at $z_{\rm c}\sim 3$
and its spin parameter is 0.05 (Barns \& Efstathiou 1987; Heavens \& Peacock 1988). 

Since the total mass of the object in our simulation is small ($10^{10}M_{\odot}$), evolution of the system and therefore the conclusion in this paper do no depend on the employed cosmology. The collapse epoch of the object in our simulation is $z_{\rm c}\sim 3$, for which the evolution is not strongly affected by the $\Lambda$ term. This is in contrast to much larger systems, such as clusters of galaxies.

The number of baryon (SPH) and dark matter particles in the spherical region is $N_{\rm SPH}=N_{\rm DM}=1005600$. 
The mass resolutions of baryon (gas and stars converted from the gas) 
and DM particles are $1.1\times 10^{3}M_{\odot}$ and $1.0\times
10^{4}M_{\odot}$, respectively. 
The gravitational softening lengths are 52 pc for baryon particles and 108 pc for DM particles.
%since those are changing comovingly before $z=10$.
The initial distribution of the particles 
is generated by COSMICS (Bertschinger 2001). 
We discuss the evolution of galaxies until $z=2$,
because the assembly history for the boundary conditions 
would not be realistic much later than the collapse epoch, $z_c$. 
But, $z\sim z_{\rm c}$ the assembly of the simulated galaxy finishes, so that the mass of the galaxy at $z=0$ would be equal to that at $z=2$.
%because the simulation volume is too small to discuss the assemble history at low-z.

The numerical technique we employ to represent the evolution of galaxies 
is a standard hybrid $N$-body/hydrodynamic code for galaxy formation.
The code includes both the radiative cooling and star formation.
However, the dynamical and radiative feedback processes
from star formation and supernova explosions are not explicitly taken into account.
%In the code, we use the Tree-algorithm with 
%a special purpose computer, GRAPE (GRAvity PipE) (Sugimoto et al. 1990), 
%for solving self-gravity, and SPH (Smoothed Particle Hydrodynamics) for ga%s dynamics. 
The length of the interaction list of each SPH particle is $N_{\rm NB}=50$. 
In the SPH simulations, the Jeans instability can be resolved correctly
for masses larger than $2 N_{\rm NB} m_{\rm SPH}$ (Bate \& Burkert 1997),
where $m_{\rm SPH}$ is mass of an SPH particle.
%%where $m_{\rm SPH}$, and $N_{\rm NB}$ are mass of an SPH particle and
%%the number of neighbor particles. 
In the simulation, we can resolve gravitational instability 
of a cloud whose mass is larger than $ 1.1\times 10^{5}M_{\odot}$.
In order to model the multiphase nature of the interstellar medium 
(e.g., Wada \& Norman 2001), we solve the energy equation with 
the radiative cooling under
$10^{4}$K and the inverse Compton cooling. 
We assume that the gas has a primordial abundance of X=0.76 
and Y=0.24 and we assume an ideal gas with $\gamma=5/3$. 
The mean molecular weight of gas, $\mu$, is set to 0.59.
%A shear-free formulation of the artificial viscosity is also implemented (Balsara 1995).

The star formation algorithm is similar to the one by Katz (1992). 
If an SPH particle satisfies all the following conditions: 
(1) the regions are in virialized halos ($\rho_{\rm SPH} > 200\rho_{\rm BG}$), 
where $\rho_{\rm BG}$ is the background density, 
(2) low temperature (T$<3\times 10^4$K), and (3) collapsing regions ($\nabla \cdot v < 0$), 
then it is converted into a collisionless star particle inheriting the velocity and the mass of the gas particle.
The local star formation rate, SFR, is assumed to be
SFR=$c_{*} m_{\rm SPH}/\tau_{\rm ff}$ with $c_{*}=1/30$, 
where $\tau_{\rm ff} = 1/\sqrt{G \rho_{SPH}}$ is a free-fall time. 

\subsection{Radiation drag model}
Next, we model the BH growth, based on the radiation drag-driven mass 
accretion. Here, we suppose a two-component system:
an inhomogeneous ISM with disk-like geometry
embedded in  a spheroidal stellar bulge. In fact, the ISM in active star-forming galaxies is highly inhomogeneous (Sanders et al. 1988; Gordon, Calzetti \& Witt 1997). Wada \& Norman (2002) also showed that the multiple SNe in a galactic center from a quasistable inhomogeneous torus around a supermassive BH.

We assume that stars and $N_{\rm c}(=10^{4})$ identical clouds are randomly distributed in a system of the bulge.\footnote{It is noted that simulations with a three times larger number of clouds did not lead to any fundamental difference for final BH mass, although at least $10^{4}$ clouds are necessary to treat the radiation transfer effect properly in clumpy ISM. The total optical depth, $\tau_{\rm bulge}$, is not also significantly affected by changing the cloud size, $r_{\rm c}$.} The optical depth of a gas cloud is $\bar{\tau}=\chi_{\rm d}\rho_{\rm gas}r_{\rm c}\simeq \chi_{\rm d}m_{\rm gas}/r_{\rm c}^{2}$, where $\rho_{\rm gas}$, $m_{\rm gas}$, and $r_{\rm gas}$ are  the densitiy, mass and the size of a cloud. The mass extinction coefficient $\chi_{\rm d}$ is given by $\chi_{\rm d}=n_{\rm d}\sigma_{\rm d}/\rho_{\rm gas}$ with the number density $n_{\rm d}$, cross-section $\sigma_{\rm d}$ and gas density $\rho_{\rm gas}$. 
In this paper, we assume $\chi_{\rm d}=300{\rm cm}^{2}{\rm g}^{-1}
\left({a_{\rm d}}/{0.1\mu {\rm m}}\right)^{-1}\left({\rho_{\rm s}}/{{\rm
g \, cm}^{-3}}\right)\left({Z}/{0.3Z_{\odot}}\right)$, 
where $a_{\rm d}$ is the grain radius, $\rho_{\rm s}$ is the 
density of solid material density within the grain (e.g., Spitzer 1978),
and $Z$ is the metallicity of gas, which are fixed at $a_{\rm d}=0.1\mu\, m$, $\rho_{\rm s}=1\,{\rm g}\,{\rm cm}^{-3}$ and $Z$=0.3$Z_\odot$.\footnote{We should keep in mind that recent observations suggest that the metallicity of the gas in the AGNs can be super-solar $Z > Z_{\odot}$ (e.g., Ohta et al. 1996; Dietrich \& Wilhelm-Erkens 2000; Maiolino et al. 2003). If this is the case, the optical depth of a gas cloud can be enhanced by a factor of 3-4.} 
The total optical depth of the bulge $\tau_{\rm bulge}$ is given by 
$\tau_{\rm bulge}=\bar{\tau}\bar{N}_{\rm int}$, where $\bar{N}_{\rm int}$ 
is the average number of clouds intersected by a light ray over a bulge scale. The $\bar{N}_{\rm int}$ is defined by $\bar{N}_{\rm int}=n_{\rm c}\pi r_{\rm c}^{2}r_{\rm bulge}=(3/4)N_{\rm c}(r_{\rm c}/r_{\rm bulge})^{2}$, where $n_{\rm c}=N_{\rm c}/\frac{4}{3}\pi r_{\rm bulge}^{3}$ is the number density of gas clouds. In this paper, we assume that the cloud covering factor is order unity, i.e. $\bar{N}_{\rm int}\approx O(1)$, according to the previous analysis. A different level of ISM clumpiness can reduce the radiation drag efficiency by a factor of 2 (Kawakatu \& Umemura 2002). We also confirm that the optical depth in a clumpy media is an order of unity, using a 3-D hydrodynamic simulations (see Fig. 1  in Wada \& Norman 2002).
%In order to test if such conditions are realistic, we have calculated the %ptical depth for an inhomogeneous distribution of the ISM suggested by 3-D%hydrodynamic simulations (see Fig. 1  in Wada \& Norman 2002). 
Even if the system is extremely gas-rich ($M_{\rm gas}=10^8 M_{\odot}$ in the central 100 pc), we found that the optical depth for the disk plane from various directions is distributed between 0.5 and 1.2, assuming the same gas/dust ratio and dust opacity in the analysis here. 
%Therefore, this implies that the assumption we employed ($\bar{N}_{\rm int%}\approx O(1)$) is valid in more realistic situation. 
%Also, the optical depth of a cloud and thus the overall optical depth of b%ulge depend on the cloud size $r_{\rm c}$. Here, $r_{\rm c}=0.01r_{\rm bul%ge}$ is assumed as a fiducial case. But, they have confirmed that no essen%tial difference in the final BH mass is found by changing $r_{\rm c}$ so a%s to enhance $\tau_{\rm bulge}$ by an order. 
Finally, using $M_{\rm gas}=N_{\rm c}m_{\rm gas}$ the total optical depth of the bulge can be re-written to be 
\begin{equation}
\tau_{\rm bulge}(t)=\bar{\tau}(t)\bar{N}_{\rm int}\simeq \frac{3\chi_{\rm d}}{4\pi}\frac{M_{\rm gas}(t)}{{r^{2}_{\rm bulge}(t)}},
\end{equation}
where $r_{\rm bulge}(t)$ and $M_{\rm gas}(t)$ are the size and the gas mass of the bulge.\footnote{In the present paper, we identify a `bulge' as a spheroidal star-forming 
region where the average number density of the gas, $n_{\rm H}$, is larger than $0.1$ cm$^{-3}$
in a spiral galaxy. This criterion corresponds to the density criterion of the star-forming region.
} 

The radiation drag, which drives the mass accretion, originates in the relativistic effect in absorption and subsequent re-emission of the radiation. This effect is naturally involved in relativistic radiation hydrodynamic equations (Umemura, Fukue, \& Mineshige 1997; Fukue, Umemura, \& Mineshige 1997). The angular momentum transfer in radiation hydrodynamics is 
given by the azimuthal equation of motion in cylindrical coordinates, 
\begin{equation}
\frac{1}{r}\frac{d(rv_{\rm \phi})}{dt}=\frac{\chi_{\rm d}}{c}[F^{\rm \phi}-
(E+P^{\rm \phi \phi})v_{\rm \phi}], \label{ldot}
\end{equation} 
where $E$ is the radiation energy density, $F^{\rm \phi}$ 
is the radiation flux, $P^{\rm \phi\phi}$ is the radiation stress tensor.
By solving radiative transfer with including dust opacity, 
we evaluate the radiative quantities, $E$, $F^{\rm \phi}$, and $P^{\rm \phi\phi}$, and thereby obtain
the total angular momentum loss rate.
% (see Kawakatu \& Umemura 2002 for the details of method). 
Then, we can estimate the mass accretion rate of the dusty ISM accreted on to a central massive dark object, $\dot{M}_{\rm MDO}$, 
by using the relation, $\dot{M}_{\rm drag}/M_{\rm gas}=-\dot{J}/J$, 
where $J$ and $M_{\rm gas}$ are the total angular momentum and gas of ISM.
In the optically-thick regime of the radiation drag,
$-M_{\rm gas}\dot{J}/{J}=L_{\rm bulge} (t)/c^{2}$, 
where $L_{\rm bulge}(t)$ is the total luminosity of the bulge.
The radiation drag efficiency depends on the optical depth $\tau_{\rm bulge}$ in proportion to $(1-e^{-\tau_{\rm bulge}(t)})$ (Umemura 2001). 
Thus, the mass accretion rate via the radiation drag is
\begin{equation}
\dot{M}_{\rm drag}=\eta_{\rm drag}\frac{L_{\rm bulge}(t)}{c^{2}}
\left(1-e^{-\tau_{\rm bulge}(t)}\right),
\end{equation}
where $L_{\rm bulge}(t)$ and $\tau_{\rm bulge}(t)$ are the total luminosity 
and the time-dependent total optical depth of the bulge.
Kawakatu \& Umemura (2002) found that 
the efficiency $\eta_{\rm drag}$ is maximally 0.34 in 
the optically thick regime. 

The radiation energy emitted by a main sequence star is $0.14\epsilon$ 
to the rest mass energy of the star, where $\epsilon$ is the energy
conversion efficiency of the nuclear fusion from hydrogen to helium, which
is 0.007. 
Thus, the luminosity of the bulge at optical and UV bands is 
simply approximated by $L_{\rm bulge}(t)\simeq 0.14\epsilon \dot{M}_{\rm
bulge}(t)c^{2}$, where $\dot{M}_{\rm bulge}(t)$ is the SFR in the bulge. 
Here, we employ a stellar initial mass function (IMF) such as
$\phi = dn/d\log m_{*}=A(m_{*}/M_{\odot})^{-\alpha}$ for a mass range of
[$m_{\rm l}$, $m_{\rm u}$], where $m_{*}$, $m_{\rm l}$, and $m_{\rm u}$
are the stellar mass, the lower mass, and the upper mass, respectively.
We assume $m_{\rm l}=0.1M_{\odot}$ and $m_{\rm u}=60M_{\odot}$,
and the index $\alpha$ is 1.35 \footnote{As for the effect of IMF, If the slope and the mass range of IMF are changed to satisfy the spectrophotometric properties of galactic bulges, then the radiation drag efficiency is altered by a factor of $\pm 50\%$ (Kawakatu \& Umemura 2004).}.
The accretion rate, equation (3), is therefore
\begin{equation}
\dot{M}_{\rm drag}\simeq 1.2\times10^{-3}\eta_{\rm drag}\dot{M}_{\rm bulge}(t)(1-e^{-\tau_{\rm bulge}(t)}), 
\end{equation}

Here we ignore the infrared luminosity from the evolved stars 
because the dust opacity for the infrared band is much smaller than 
that for the optical and UV bands.
$\dot{M}_{\rm bulge}$ and $\tau(t)$ are directly given from the
numerical simulations.
%and thus we neglect the contribution from them. 
%Thus, substituting the physical quantities (SFR and optical depth) of 
%bulge components obtained by our numerical simulations for equation (3), 
%we can estimate the mass accretion rate due to the radiation drag.
The total mass of dusty ISM 
accreted to the central massive dark object (MDO), $M_{\rm MDO}(t)$, 
is obtained by 
\begin{equation}
M_{\rm MDO}(t)=\int_{0}^{t}\dot{M}_{\rm drag}dt. 
\end{equation}

As seen in equation (3), (4) and (5), the linear relation between the MDO mass and the bulge mass is a direct result of the radiation drag mechanism. The possible mass accreted by the radiation drag in the optically thick limit is given by 
\begin{equation}
M_{\rm MDO, max}=\eta_{\rm drag}\int_{t_{0}}^{t_{1}}L_{\rm bulge}/c^{2}\, dt \simeq 5\times 10^{-3}M_{\rm bulge}, 
\end{equation}
where $t_{0}$ is 0.1 Gyr ($z\sim 25$) which corresponds to the epoch we detect the progenitor of galaxy firstly, and $t_{0}$ is 2.6 Gyr ($z\sim 2$). 
The correspondance between time and redshift is based on the cosmological model we adopted.

In this model, we should distinguish the BH mass from that of an MDO, 
although the mass of an MDO is often regarded as BH mass 
from an observational point of view. The radiation drag is not likely to
remove the angular momentum thoroughly, and thus some residual angular
momentum will terminate the radial contraction of the accreted gas (Sato
et al. 2004). Hence, the dusty ISM probably 
forms a compact rotating torus. 
In this nuclear torus, we suppose that the mass accretion 
onto the BH horizon is determined by the Eddington rate,
and that 
the BH mass grows according to 
\begin{equation}
M_{\rm BH}(t) =M_{0}e^{t/t_{\rm Edd}}, 
\end{equation}
where $t_{\rm Edd}$ is the Eddington time scale, 
$t_{\rm Edd}=\eta_{\rm BH}M_{\rm BH}c^{2}/L_{\rm Edd}$, with
the energy conversion efficiency, $\eta_{\rm BH}$, 
and the Eddington luminosity, $L_{\rm Edd}$.
Unless otherwise stated, $\eta_{\rm BH}$ is assumed to be 0.42, which is 
the conversion efficiency of an extreme Kerr BH.
Recently, Shibata (2004) has found that a rigidly rotating very massive star (VMS)
with several $100M_{\odot}$ can be unstable for a softer equation of
state, 
and eventually it forms a BH. 
In addition, the theory of stellar evolution reveals that
the nuclear burning in VMSs above $260M_{\odot}$ is unable to halt 
gravitational collapse (e.g., Heger et al. 2003). Thus, the VMSs inevitably evolve into massive
BHs without supernova explosions. 
Here, we assume $260M_{\odot}$ as the mass of the seed black hole $M_{0}$.

\section{Results}
On the basis of the coevolution model described in the previous section,
we estimate the mass accretion driven by the radiation drag during 
hierarchical galaxy formation. Next, we reveal the relationship
between the growth of a BH and that of a dark matter halo. 
Finally, we discuss the relation between AGN activity 
and the properties of the bulges.
% and then we suggest a new tool to
%search the BH growing objects 
%at high redshift.

\subsection{Mass accretion rate via radiation drag}
Evolution of the SFR and the optical depth of bulge ($\sim$ 1 kpc) 
are shown in Figure 1. 
Before $z\sim 4$, the SFR and the optical depth increase, 
while they decrease rapidly for $z<4$. 
This can be understood as follows: 
At high-$z$ ($z>4$), supply of the gas due to mergers of smaller 
proto-galaxies and consumption of the gas in the central part of
the galaxy (bulge) are almost balanced.
Thus, both the SFR and the optical depth are enhanced.
At low-$z$ ($z<4$), accretion of the gas to the bulge associated with merger events 
is decreased. 
The gas in the bulge is consumed by the star formation.
As a result, it makes the gas of bulge poor.
As seen in Figure 1, we found that the evolution of the SFR 
and the optical depth are not smooth, but episodic. This episodic
growth corresponds to the phase of the high mass accretion onto
the bulge component triggered by major mergers with some time delays,
which are typically $10^{7}{\rm yrs}$ (see Saitoh \& Wada 2004 in
details).

%\newpage
%\begin{figure}
%\plotone{f1.eps}
\vspace{5mm}
\epsfxsize=8cm 
\epsfbox{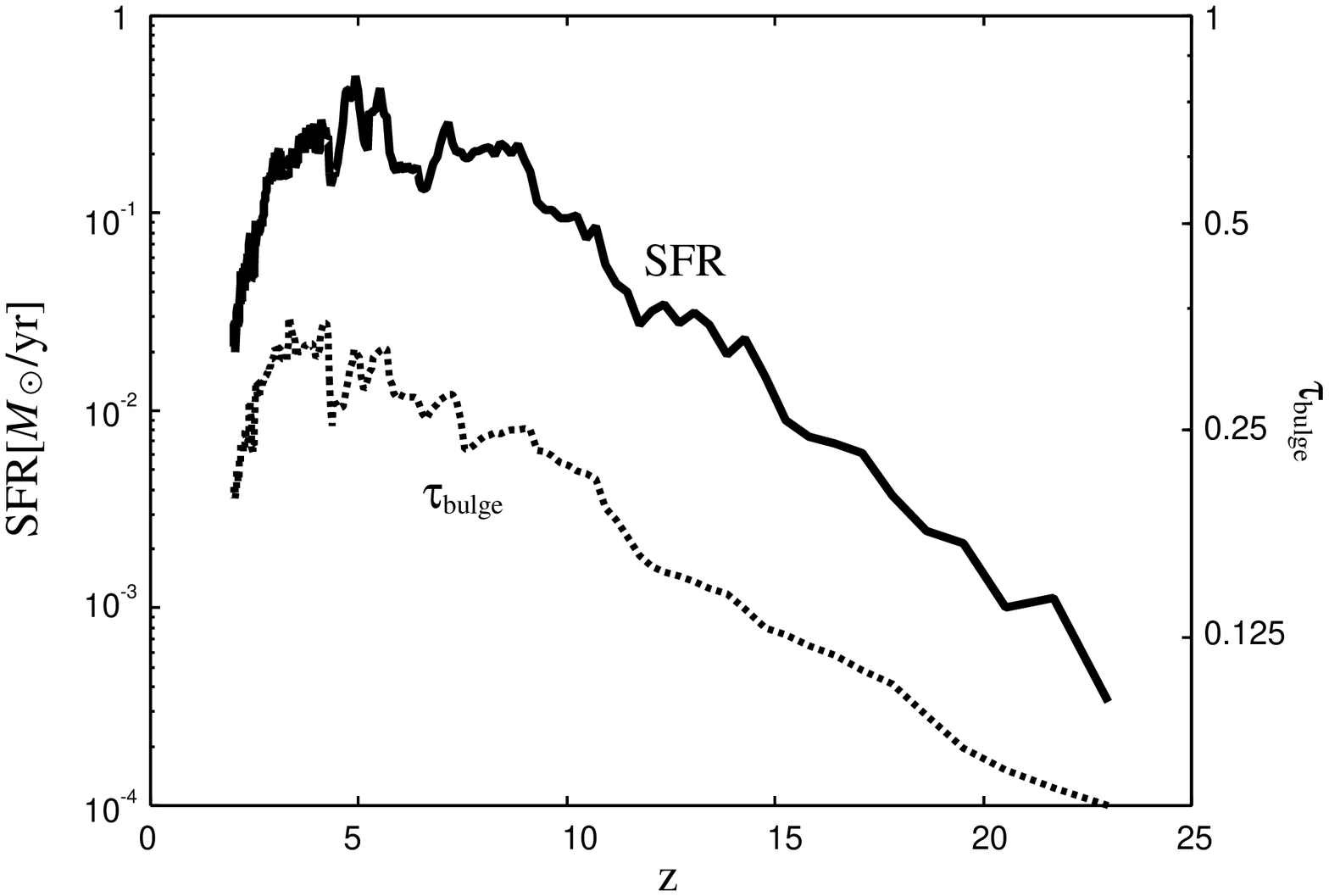}
\figcaption
{
Redshift evolution of the star formation rate (in units of $M_{\odot}$ yr$^{-1}$)
 and total optical depth ($\tau_{\rm bulge}$) of the bulge at redshift ($z$). At $z>4$,
both the SFR and the optical depth increase with time, while they
decrease at low-$z$ ($z<4$). 
%The evolution of the SFR and the optical depth are not smooth, but 
%episodic, with a time-scale of typically $10^{7}{\rm yrs}$.
%These episodic changes are triggered by major mergers.
}
%%\label{Figure 1}
%\end{figure}
\vspace{2mm}

Figure 2 is evolution of the mass accretion rate due to the radiation drag 
($\dot{M}_{\rm drag}$), the rate of mass accretion onto a BH
($\dot{M}_{\rm BH}$), and the Eddington mass accretion rate
($\dot{M}_{\rm Edd}$). 
It is clear that 
$\dot{M}_{\rm drag}$ is also episodic, reflecting the evolution of 
the SFR and optical depth (Figure 1 and eq.$[4]$).
We have also found that the averaged mass accretion rate is
$\approx 10^{-5}M_{\odot}$ yr$^{-1}$. This rate is comparable 
to the Eddington mass accretion rate for a black hole mass with
$10^{4}M_{\odot}$, that is , 

\begin{equation}
\dot{M}_{\rm Edd}=\frac{1}{\eta_{\rm BH}}\frac{L_{\rm Edd}}{c^{2}}
\approx 10^{-5}M_{\odot}{\rm yr}^{-1}\left(\frac{\eta_{\rm
BH}}{0.42}\right)^{-1}\left(\frac{M_{\rm BH}}{10^{4}M_{\odot}}\right).
\end{equation} 

In Figure 2, the Eddington mass accretion rate (eq. $[8]$) 
is larger than $\dot{M}_{\rm drag}$ after $z \sim 5$. 
Since the BH mass equals the mass of MDO at $z\approx 4$ (see Figure 3),
the mass accretion onto the BHs after $z\sim 4$ would be controlled
by the mass accretion to the MDO via the radiation drag, 
i.e. $\dot{M}_{\rm BH} = \dot{M}_{\rm drag}$.
Figure 2 shows that 
$\dot{M}_{\rm BH} > \dot{M}_{\rm drag}$ in a period of $4.2 < z < 4.8$ 
(the shaded area in Figure 2). In this paper, we call this period a
`BH-growing phase',  which is $\approx 10^{8}{\rm yr}$. 

%\newpage
%\begin{figure}
%\plotone{f2.eps}
\vspace{5mm}
\epsfxsize=8cm 
\epsfbox{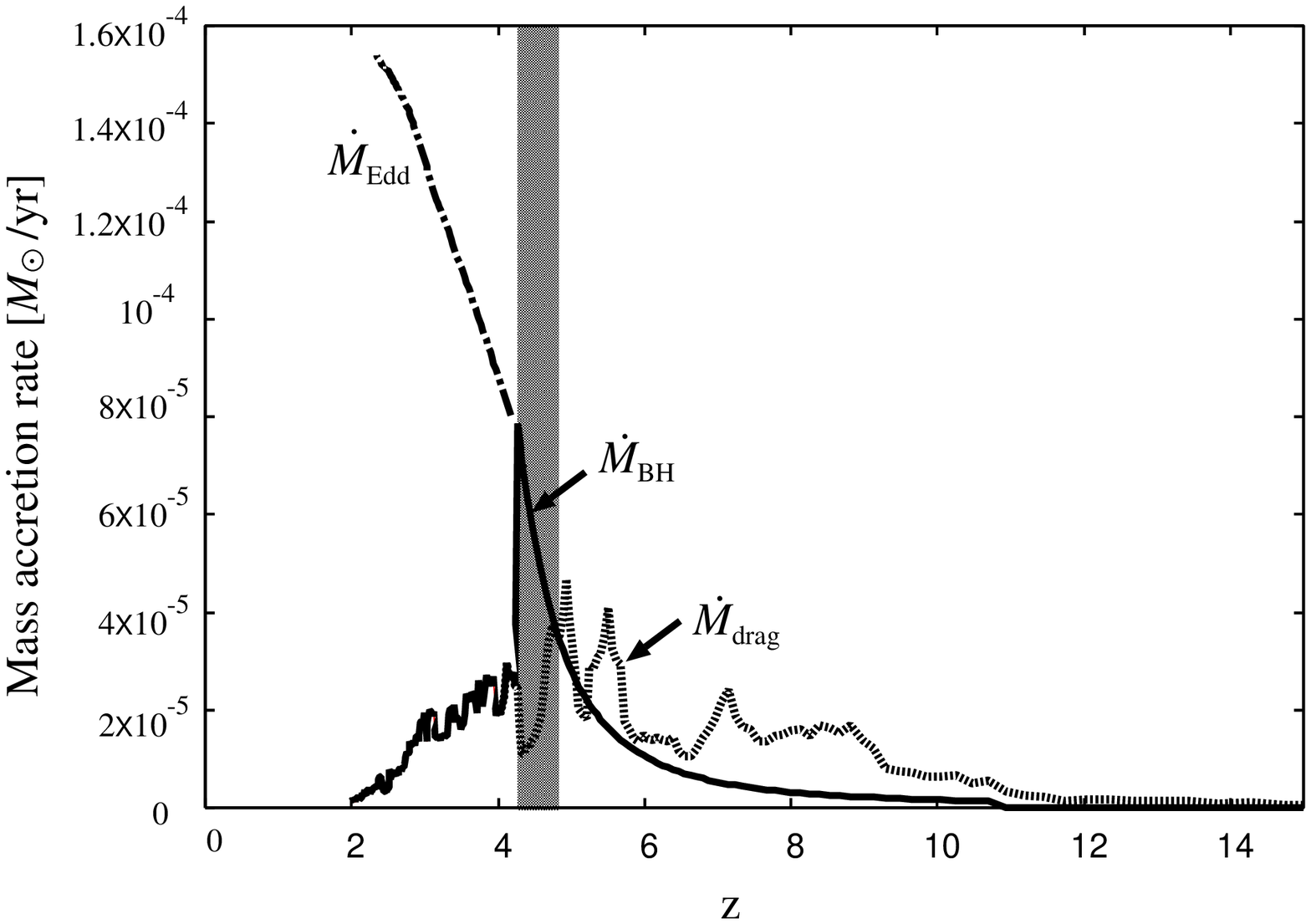}
\figcaption
{Same as Fig. 1, but for the mass accretion rate due to the radiation drag ($\dot{M}_{{\rm drag}}$), 
mass accretion rate onto a BH ($\dot{M}_{\rm BH}$), and the Eddington
mass accretion rate ($\dot{M}_{\rm Edd}$).
The averaged mass accretion rate due to the radiation drag is a few $10^{-5}M_{\odot}{\rm yr}^{-1}$, 
which is comparable to the Eddington mass accretion rate for a black
hole mass with $\approx 10^{4}M_{\odot}$. 
After $z\simeq5$, $\dot{M}_{\rm Edd}$ exceeds $\dot{M}_{{\rm drag}}$.
For $z < 4$, $\dot{M}_{\rm BH}$  follows $\dot{M}_{{\rm
drag}}$ because the mass of the BH reaches that of MDO at $z\approx
4$ (see also Figure 3). The shaded area ($4.2 < z < 4.8$) corresponds to the BH growing
phase ($\dot{M}_{\rm BH} > \dot{M}_{\rm drag}$).
}
%%\label{Figure 2}
%\end{figure}

\subsection{Coevolution of MBHs, bulges and dark matter halos}
Figure 3 shows that evolution of 
masses of the dark matter ($M_{\rm halo}$), stellar component in the bulge ($M_{\rm bulge}$), 
the gas in the bulge ($M_{\rm gas}$), MDO ($M_{\rm MDO}$), and the
massive BH ($M_{\rm BH}$). $M_{\rm halo}, M_{\rm bulge} $, and $M_{\rm
gas}$ are directly 
obtained from the numerical simulation of galaxy formation (\S 2.1), and
we obtain $M_{\rm MDO}$ and $M_{\rm BH}$ from $\dot{M}_{\rm drag}$ and $\dot{M}_{\rm BH}$ (Figure 2).

The BH mass reaches $M_{\rm MDO}$ at $z\approx 4$. 
As seen in Figure 3, during $z > 4$, the MBH grows due to the Eddington mass accretion. 
At $z < 4$, the growth rate of the MDO is reduced because $\dot{M}_{\rm
drag}$
becomes smaller than $\dot{M}_{\rm Edd}$, and $\dot{M}_{\rm drag}$ 
declines owing to the decrease in the SFR and the optical depth of bulges (Figure 1).
%After $z\sim 4$, the MDO is identical to the MBH, and its growth is determined by
%the gas accretion due to the radiation drag.
As a result, we find that an intermediate MBH with $\approx 3\times 10^{4}M_{\odot}$ 
can be formed at $z\sim 2$ in a small spiral galaxy (with a total mass of $10^{10} M_\odot$).
At $z=2$, the growth of the simulated galaxy finishes, 
so that the BH growth also would stop after $z=2$ because of the lack of gas in the bulge.
In Figure 3, it is also found that the significantly massive dusty torus with $M_{\rm MDO} > M_{\rm BH}$ exists in the early phase of the BH growth ($z\approx 5$). The massive torus itself produced by the starburst in bulges would obscure the nucleus in the edge-on view and make a type 2 nucleus. 
Therefore, it is suggested that not only a nuclear starburst in a torus around a supermassive BH like a MDO (Ohsuga \& Umemura 1999; Wada \& Norman 2002; Watabe \& Umemura 2005), but also a starburst in a bulge may contribute to the AGN obscuration in the BH growing objects.

%\newpage
%\begin{figure}
%\plotone{f3.eps}
\vspace{5mm}
\epsfxsize=8cm 
\epsfbox{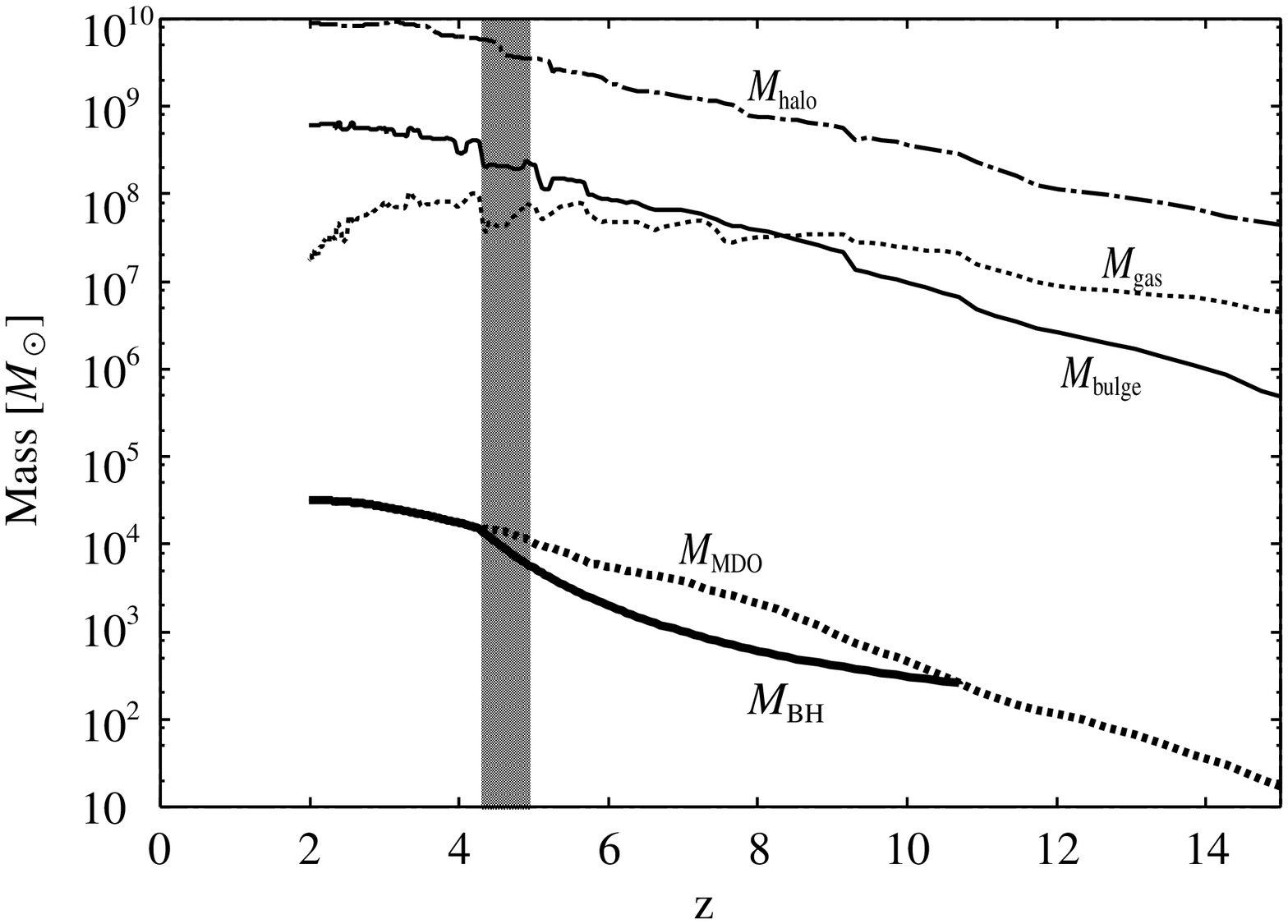}
\figcaption
{Same as Fig. 1, but for masses of the dark halo ($M_{\rm halo}$) , gas ($M_{\rm gas}$), 
and bulge ($M_{\rm bulge}$). 
Evolution of the mass of black hole (BH) and the massive dark object (MDO) are also
plotted. The mass of the seed black hole is assumed to be
 $M_{0}=260M_{\odot}$ (see eq.$[7]$). 
%The ordinate is mass in units of solar mass. 
%The mass of the dark matter halo is $M_{\rm halo}$. 
%The mass of the stellar component in the bulge and that of the gas in the bulge are 
$M_{\rm bulge}$ and $M_{\rm gas}$, respectively, while $M_{\rm MDO}$ is the mass of MDO 
and $M_{\rm BH}$ is the mass of the massive BH. 
It shows that the MDO mass is proportional to the bulge mass.
The BH mass reaches the MDO mass at $z\approx 4$.
}
%\label{Figure 3}
%\end{figure}
\vspace{2mm}
%\newpage
%\begin{figure}
%\plotone{f4.eps}
\vspace{2mm}
\epsfxsize=8cm 
\epsfbox{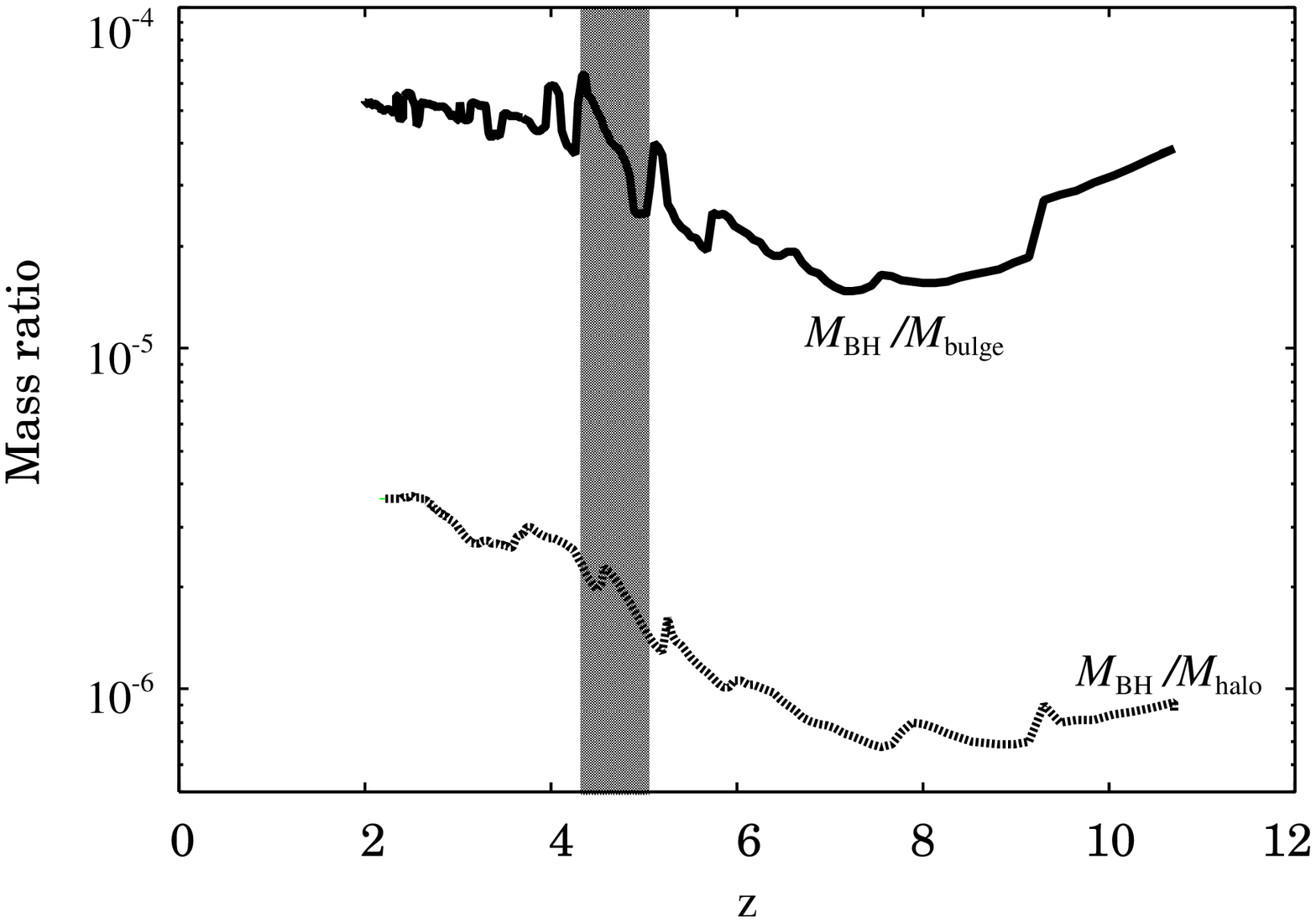}
\figcaption
{
Evolution of the mass ratios, $M_{\rm BH}/M_{\rm bulge}$ and 
$M_{\rm BH}/M_{\rm halo}$, against the redshift. 
The mass ratios are nearly constant within a factor of 2-3, but 
before $z > 8$, growth of the BH is slower than that of the bulge. 
Note that the mass ratios increase more rapidly during the BH growing phase (the 
shaded area which lasts $\approx 10^{8}{\rm yr}$). 
After $z\sim 4$, the mass ratios are almost constant with 
$M_{\rm BH}/M_{\rm bulge}\approx 5\times 10^{-5}$ and $M_{\rm BH}/M_{\rm halo}\approx3\times 10^{-6}$. 
%But, the evolutions of these mass ratios are not smooth, but step function%-like which cause the scatter by a factor of 2-3.
}
%\label{Figure 4}
%\end{figure}
\vspace{2mm}

Figure 4 shows evolution of mass ratios between the BH and the bulge or the halo,
 i.e. $M_{\rm BH}/M_{\rm bulge}$ and $M_{\rm BH}/M_{\rm halo}$. 
We find that before $z\approx 8$, the growth of the bulge and dark matter halo
are faster than that of the BH. 
After $z\approx 8$, the mass ratios increase with time because 
the growth of the BH is dominated by the Eddington mass accretion rate, 
and the rate is increasing exponentially (see Figure 2).
One should note, however, that the mass ratios change by a factor of 2-3 
in the BH-growing phase ($\approx 10^{8}{\rm yr}$). 
This change is caused by a time-lag between 
the BH growth and the growth of the bulge or the halo.  
If this is also the case in real galaxies, the scatter in
the observed $M_{\rm BH}-M_{\rm bulge}$ relation may
be because of this time-lag.
%Theese scatter would be caused by a time-lag between 
%the BH growth and the growth of bulge (halo).  
On the other hand, the mass of the MDO equals that of the BH ($z < 4$), 
and the mass ratios do not significantly change 
with $M_{\rm BH}/M_{\rm bulge}\approx 5\times 10^{-5}$ 
and $M_{\rm BH}/M_{\rm halo}\approx 3\times 10^{-6}$.
%Therefore, we would predict that the scatter of the observed scaling relation in the BH growing objects is larger than that in nearby normal galaxies. 
Therefore, we would predict that the scatter in the scaling relation
is larger in the BH-growing objects at high-$z$ than in nearby well-evolved galaxies. 

%However, the evolution of $M_{\rm BH}-M_{\rm bulge}$ and $M_{\rm BH}-M_{\rm halo}$ 
%relation is step function like, which cause the variation by a factor of
%2-3, because the time lag between the BH growth and the bulge (halo)
%growth exists. 
%Thus, this features would be a reason for the scatter for the observed scaling relations.  

In Figure 5, we plot mass of the MDO and the BH against the halo mass.
This reveals that the MDO and the BH coevolve with the dark matter halo 
from $z\sim 10$ to $z\sim 2$. 
The masses of the MDO and the BH increase with the development of the dark halo. 
The mass ratio of the BH to the halo ($M_{\rm BH}/M_{\rm DM}$) increases
gradually from $10^{-6}$ to $3\times 10^{-6}$ 
from $z\sim 7$ to $z\sim 2$ (see also Figure 4).
%[KW: The following paragraph should be omitted]
%The coevolution of the BH growth and the dark halo growth result 
%from the following process: 
%1) First, the dark matter halo grows as a result of the mergers of 
%a small proto-galaxy. 2) Afterward, by the transfer of the angular
%momentum due to the gravitational torque the gas accretes to a bulge
%region ($R \sim $ several 100pc).
%Then, the starburst within the bulge is triggered. 
%3) Because of this starburst, both the radiation energy density 
%and the optical depth increase in the bulge component of a small spiral
%galaxy. 
%4) As a result, the radiation drag works efficiently, 
%and it removes the angular momentum from the gas. 
%5) Finally, the gas assembled in the galactic center accretes onto the seed BH. 

From these arguments, the $M_{\rm BH}$-$M_{\rm halo}$ correlation indicates 
that variation of the dark matter halo potential associated with merging
processes 
positively links with the mass accretion toward the galactic center 
via the radiation drag.
Our model suggests that a BH mass is mutually related to the mass of 
a bulge and that of a dark matter halo throughout the history of the
galaxy formation. 
Comparison with observations is discussed in \S 4.

%\newpage
%\begin{figure}
%\plotone{f5.eps}
\vspace{5mm}
\epsfxsize=8cm 
\epsfbox{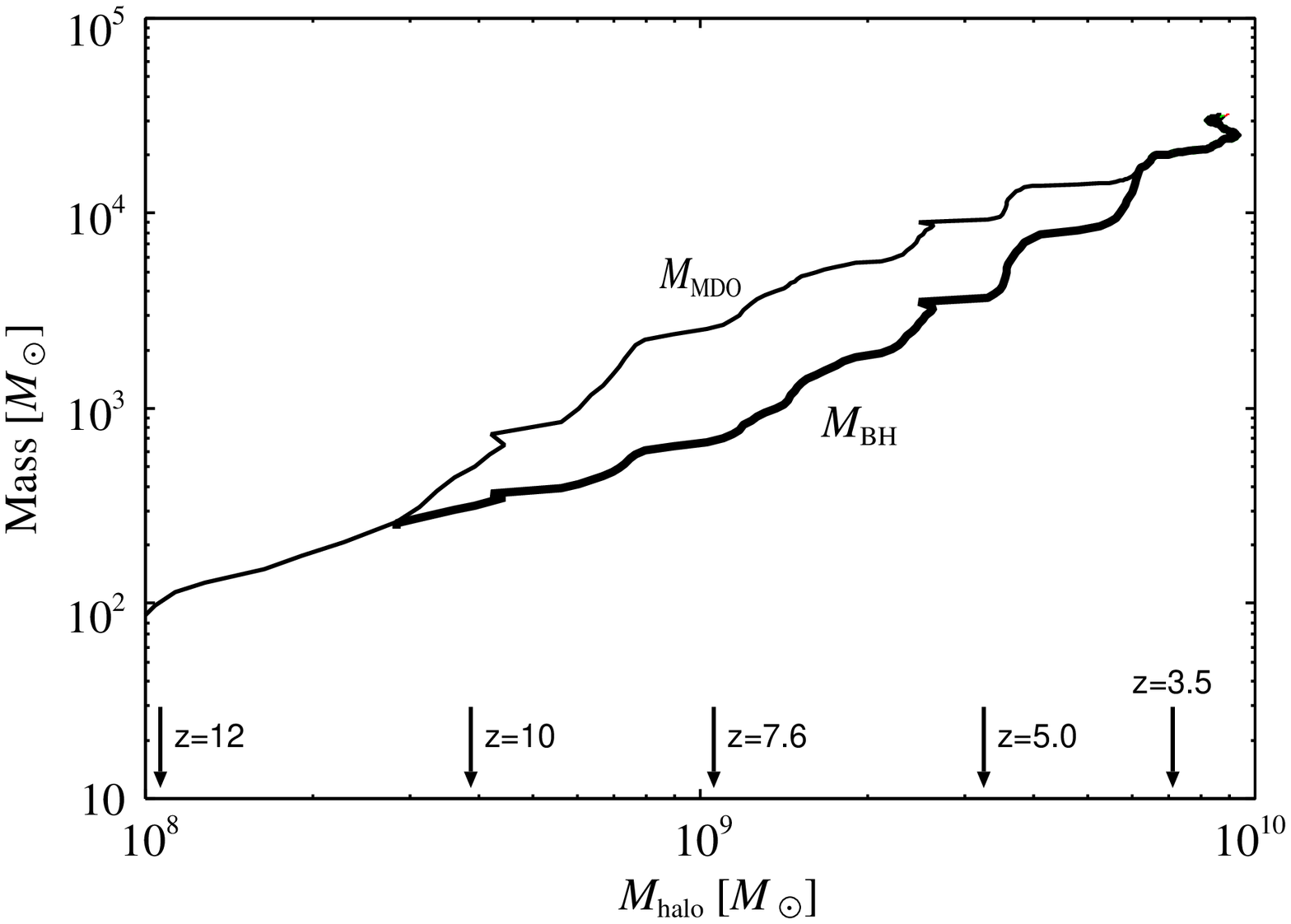}
\figcaption
{Masses of the BH (thick line) and MDO (thin line) 
are plotted against mass of the dark halo.
The arrows indicate the masses of the dark halo at labeled redshifts. 
It shows that the BH coevolves with the dark matter halo from $z\sim 10$ to $z\sim 2$. }
%The mass ratio of the BH to the dark halo is nearly constant with
%$M_{\rm BH}/M_{\rm halo}\approx 10^{-6}$ (see also Fig. 4).}
%\label{Figure 5}
%\end{figure}
\vspace{2mm}

\subsection{AGN activity-host relation}
In this section, we examine the relation between the AGN activities and  
the properties of bulge components. 
Evolution of the bulge luminosity ($L_{\rm bulge}$) at optical 
and UV-band and the AGN luminosity ($L_{\rm AGN}$) are plotted in Figure 6. 
During $z>4$, the AGN luminosity increases with the time,
because the mass accretion is determined 
by the Eddington rate.
After $z\sim 4$, the AGN luminosity is limited by $\dot{M}_{\rm drag}$
(see Figure 3). Thus, the AGN luminosity
exhibits a peak around $z\sim 4$, when $M_{\rm MDO} \sim M_{\rm BH}$. 
 As seen in this figure, the AGN luminosity is always 
smaller than the bulge luminosity; in other words, no quasar phase, 
i.e. AGN luminosity dominant phase, appears. However, the luminosity ratio of
the AGN to the bulge exhibits the maximal value ($L_{\rm AGN}/L_{\rm
bulge}\approx 0.1$) at $z\sim 4$.
% because of $L_{\rm AGN}\approx 5\times
%10^{8}L_{\odot}$ and 
%$L_{\rm bulge}\approx 5\times 10^{9}L_{\odot}$ at this redshift.
This suggests that some small spiral galaxies at high-$z$ 
could show the same level of typical low luminosity AGNs in the local universe.

Figure 7 shows the relation between the X-ray luminosity, $L_{\rm X}$,
of the AGN and the CO luminosity, $L_{\rm CO}$, of the bulge. 
Here, the X-ray luminosity $L_{\rm X}$ is estimated assuming
 $L_{\rm X}=\epsilon_{\rm X}L_{\rm AGN}$,
where $\epsilon_{\rm X}$ is the X-ray emitting efficiency, which is
supposed to be 0.1. The CO luminosity is derived from the gaseous mass 
assuming a conversion factor, $X_{\rm CO}=M_{\rm gas}/L_{\rm CO}=
4.6M_{\odot}/{\rm K}\,{\rm km}{\rm s}^{-1}\,{\rm pc}^{2}$
(de Breuck et al. 2003). 
%Although the X-factor could affect the CO abundance (Sakamoto
%1996) and Arimoto et al. (1996) found smaller $X_{\rm CO}$ in the
%central regions, in this paper by assuming $4.6M_{\odot}/{\rm K}\,{\rm
%km}{\rm s}^{-1}\,{\rm pc}^{2}$, the CO luminosity of the bulge is given
%by $L_{\rm CO}=M_{\rm gas}/X_{\rm CO}$. 
%As a result,
We found that the X-ray luminosity is positively linked 
with the CO luminosity for a wide range of luminosities, i.e.
$L_{\rm X} \sim 10^{38} ({\rm ergs} \; {\rm s}^{-1}) (L_{\rm CO}/{\rm K}\, {\rm km}\, {\rm s}^{-1} \, {\rm kpc}^2)^2$.
%This result shows that the AGN activities have a strong effect on the 
%gaseous mass of the bulge. 

%[KW: I do not understand this paragraph.]
In Figure 7, we also find that some points obviously 
deviate from the linear relation. These points correspond 
to the BH-growing phase ($4.2< z < 4.8$; the shaded area in Figure 2, 3, 4 and 6). 
This deviation implies that the growth of the BH via 
the Eddington mass accretion is much faster than 
that of gas mass in the bulge in this phase (see Figure 3).
In other words, except for the BH-growing phase, our result shows that 
the time scale of the BH growth is comparable to that of the growth of the gas 
component in the bulge.  
%This phase might correspond to the BH growing objects 
%like narrow line Seyfert 1s (NLS1s). 
%NLS1s are considered to be type 1 Seyfert galaxies 
%in the early stage of their evolution (Mathur 2000).
Hence, by using the deviation from the linear relation for $L_{\rm X}-L_{\rm CO}$ diagram, we may look for the BH growing objects at high-$z$ that we have missed so far.

%\newpage
%\begin{figure}
%\plotone{f6.eps}
\vspace{5mm}
\epsfxsize=8cm 
\epsfbox{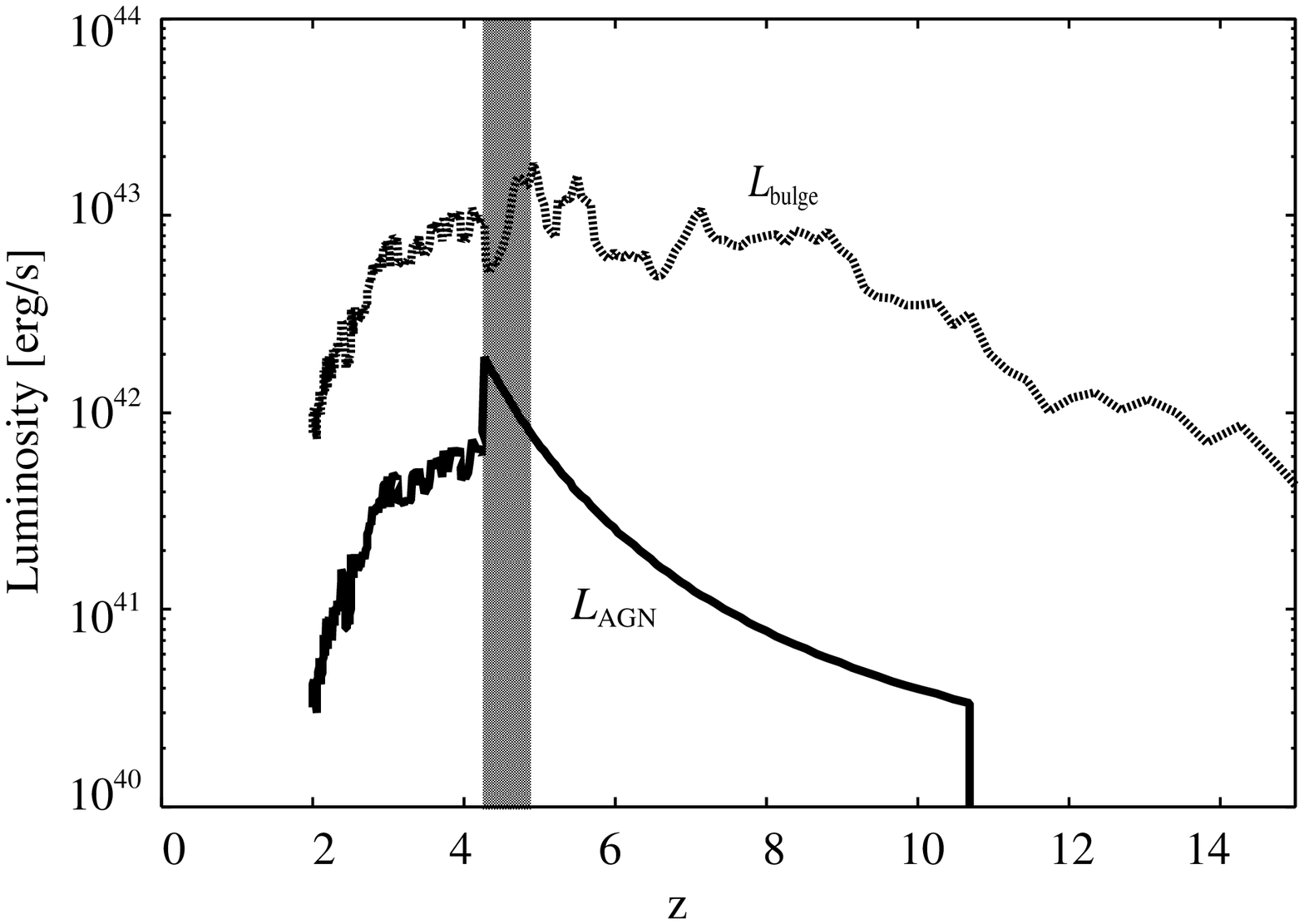}
\figcaption
{
AGN and bulge luminosity as a function of redshift. 
%The ordinate is the luminosity in units of solar luminosity.
Here, we assume that $L_{\rm AGN}$ is the Eddington luminosity before $z\sim 4$. 
After $z\sim 4$, the AGN luminosity is determined by the mass accretion
due to the radiation drag.
After the AGN luminosity exhibits a peak around $z\sim 4$, it fades out. 
At $z\approx 4$, the luminosity ratio of the AGN to the bulge exhibits
the maximal value ($L_{\rm AGN}/L_{\rm bulge}\approx 0.1$).}
% and thus it corresponds to the typical low luminosity AGNs.}
%\label{Figure 6}
%\end{figure}

\section{Comparison with observations}
\subsection{Massive BHs in Lyman break galaxies}
In the present model, we have revealed that a small spiral galaxy 
(with a total mass of $10^{10}M_{\odot}$) can have an IMBH with $\approx
10^{4}-10^{5}M_{\odot}$ at $z \sim 2$. This is consistent with the mass range
of the massive BH in dwarf galaxies with AGNs in the local universe
(Barth et al. 2004; Greene \& Ho 2004; Barth, Greene, \& Ho 2004). 
Moreover, according to Figure 6, the AGN luminosity exhibits 
a peak around $z\approx 4$ (BH-growing phase). This phase has several
characteristic properties; (1) The X-ray luminosity of the AGN is relatively
low $L_{\rm X}\approx 5\times 10^{7}L_{\odot}$ if $L_{\rm X}=0.1L_{\rm
AGN}$. (2) The luminosity of the bulge at optical and UV-band is $5\times
10^{9}L_{\odot}$. (3) The SFR is $\approx 0.3M_{\odot}\,{\rm yr}^{-1}$. (4)
The stellar mass is $\approx 10^{8}M_{\odot}$. (5) An IMBH
of $\approx 10^{4}M_{\odot}$ exists in the bulge. (6) The total optical
depth of the bulge is $\tau_{\rm bulge}\approx 0.4$.

As for the host galaxy, it has recently been suggested that Lyman break galaxies 
(LBGs), which are starburst galaxies at high-$z$, 
may be in the forming phase of a galactic bulge 
(Friaca \& Terlevich 1999; Matteucci \& Pipino 2002).
% These LBGs are
%widely known as high-z starburst galaxies, and 
The SFR in the LBGs is $\sim 3-300
M_{\odot} {\rm yr}^{-1}$. LBGs have a typical luminosity $\sim
10^{10}-10^{11}L_{\odot}$, a stellar mass $\sim 10^{10}-10^{11}M_{\odot}$, 
and they are observed to be optically thin (e.g., Shapley et al. 2001). 
They also exhibit strong clustering at $z$$\sim 3$, 
suggesting that hierarchical clustering is on-going. 
In addition, the Chandra
X-ray observatory has detected the hard X-ray of LBGs at $z=2-4$ with the
luminosity of $\sim 10^{8}L_{\odot}$ 
although it is still uncertain whether the X-ray emission arises from
AGNs (Brandt et al. 2001). 
Comparing these properties of LBGs with our predictions (i.e. (1)-(6)), the small spiral galaxy (with a total mass of
$10^{10}M_{\odot}$) in the BH-growing phase
may correspond to low-mass counterparts of 
the LBGs.
Extrapolating the scaling relation that we found to 
the observed LBGs, they would have the MBHs with
$\approx 10^{6}-10^{7}M_{\odot}$.
According to recent observations, only
$3\%$ of LBGs show AGN activity in the rest-frame hard X-ray band
(Namdra et al. 2002) and optical band (Shapley et al. 2001). 
However, it has not been clear that this $3\%$ fraction reflects the
duty cycle of mass accretion to BHs (see Shapley et al. 2001 in details)
or the possibility that the LBGs have massive BHs. 
Hosokawa (2004) claimed, assuming $M_{\rm BH}/M_{\rm bulge}\approx 0.001$,
that $\sim 10\%$ of LBGs at $z\sim 3$ can have a massive BH with $\sim
10^{7}M_{\odot}$ to reproduce the local mass function of
SMBHs (Salucci et al. 1999; Yu \& Tremaine 2002; Aller \& Richstone
2002; Shankar et al. 2004). 
%Therefore, from both theoretical and statistical viewpoints, 
%some LBGs at $z\sim 3$ could have massive BHs with $\sim 10^{7}M_{\odot}$.

\subsection{$M_{\rm BH}-M_{\rm bulge}$ and $M_{\rm BH}-M_{\rm halo}$ relations}
Barth et al. (2004) suggested that the correlation between BH mass and 
stellar velocity dispersion (the $M_{\rm BH}-\sigma$ relation) 
holds on a mass scale of an intermediate BH with $\approx
10^{4}-10^{6}M_{\odot}$. In addition, Baes et al. (2003)
found a correlation between the BH mass and halo mass, i.e. $M_{\rm
BH}/10^{8}M_{\odot} \sim 0.11(M_{\rm halo}/10^{12}M_{\odot})^{1.27}$, by
using the $M_{\rm BH}-\sigma$ relation. 
This relation gives  $M_{\rm BH} = 
3\times 10^{4}M_{\odot}$ for $M_{\rm halo}=10^{10} M_\odot$,
which is comparable to
our prediction ($M_{\rm BH}\approx 3\times 10^{4}M_{\odot}$ at $z\sim 2$).
However, it should be noted that there are 
large uncertainties in the empirical laws, as mentioned by Ferrarese (2002).

In our model, the final BH mass-to-bulge mass ratio is $\approx 5\times 10^{-5}$, which is much smaller than the observed value ($\approx 10^{-3}$) in nearby large galaxies. Thus, if our scenatio is correct, it is expected that small spiral galaxies at high-$z$, which are not direct counter parts of the dwarf galaxies at low-$z$, could have the IMBHs and the smaller BH mass-to-bulge mass ratios.
Currently, it is difficult to detect the IMBHs in small galaxies at high-$z$, and therefore we can not directly prove our prediction, namely the 
small BH mass-to-bulge mass ratio. Moreover, there is also a room in the
theoretical model, especially on effects of mechanical, radiative
and chemical feedback processes from star formation. For instance,
it is not trivial whether the stellar feedback affects on the SFR positively or negatively.  Therefore, it is ultimately necessary to perform 
high-resolution radiative hydrodynamical simulations for galaxy
formation taking into account these effects explicitly. 

In this paper, we focus on the formation of a MBH due to the radiation drag 
in a small spiral galaxy. 
However, the process discussed here could be applied to more massive galaxies 
because the transfer of the angular momentum via the radiation drag 
is independent of the mass scale of the galaxies. 
By using the BH mass-to-halo mass relation for a small spiral 
galaxy ($M_{\rm BH}/M_{\rm halo}\approx 10^{-6}$),  
we can predict that the massive spiral galaxies with $M_{\rm halo}=10^{12}M_{\odot}$ 
have MBHs with $10^{6}-10^{7}M_{\odot}$. In addition, 
the observations have suggested the SFRs of the massive galaxies at high-$z$ are
10-100 times higher than those of small galaxies like the one considered here.
Thus, the final BHs might achieve $\approx 10^{8}M_{\odot}$ because the effect
of the radiation drag is linearly proportional to the SFR (eq.$[4]$).
If this is the case, host galaxies of the luminous quasars at high redshift 
would be star-forming or post-starburst galaxies.
%Therefore, this object may corresponds 
%to a powerful AGN like a quasar.
On the other hand, we should note that the mass of a BH would depend 
on the morphology of the galaxies even if the mass of the dark halo is the same. 
This is because the rate of mass accretion onto the BHs is not
determined by the disk components, but by the bulge components in the host
galaxies, due to the effects of geometrical dilution and opacity (see Kawakatu \& Umemura 2004 in details). Therefore, the morphology
differences of host galaxies would cause the large scatter in the mass
relation of the BHs to the halos in spiral galaxies. 
In fact, some authors claim that the $M_{\rm BH}-M_{\rm halo}$ relation 
is much weaker than the $M_{\rm BH}-M_{\rm bulge}$ relation in spiral galaxies 
(Salucci et al. 2000; Zasov et al. 2004). 
%These observational results may be explained by the effect of morphology
%in spiral galaxies. 

%\newpage
%\begin{figure}
%\plotone{f7.eps}
\vspace{5mm}
\epsfxsize=8cm 
\epsfbox{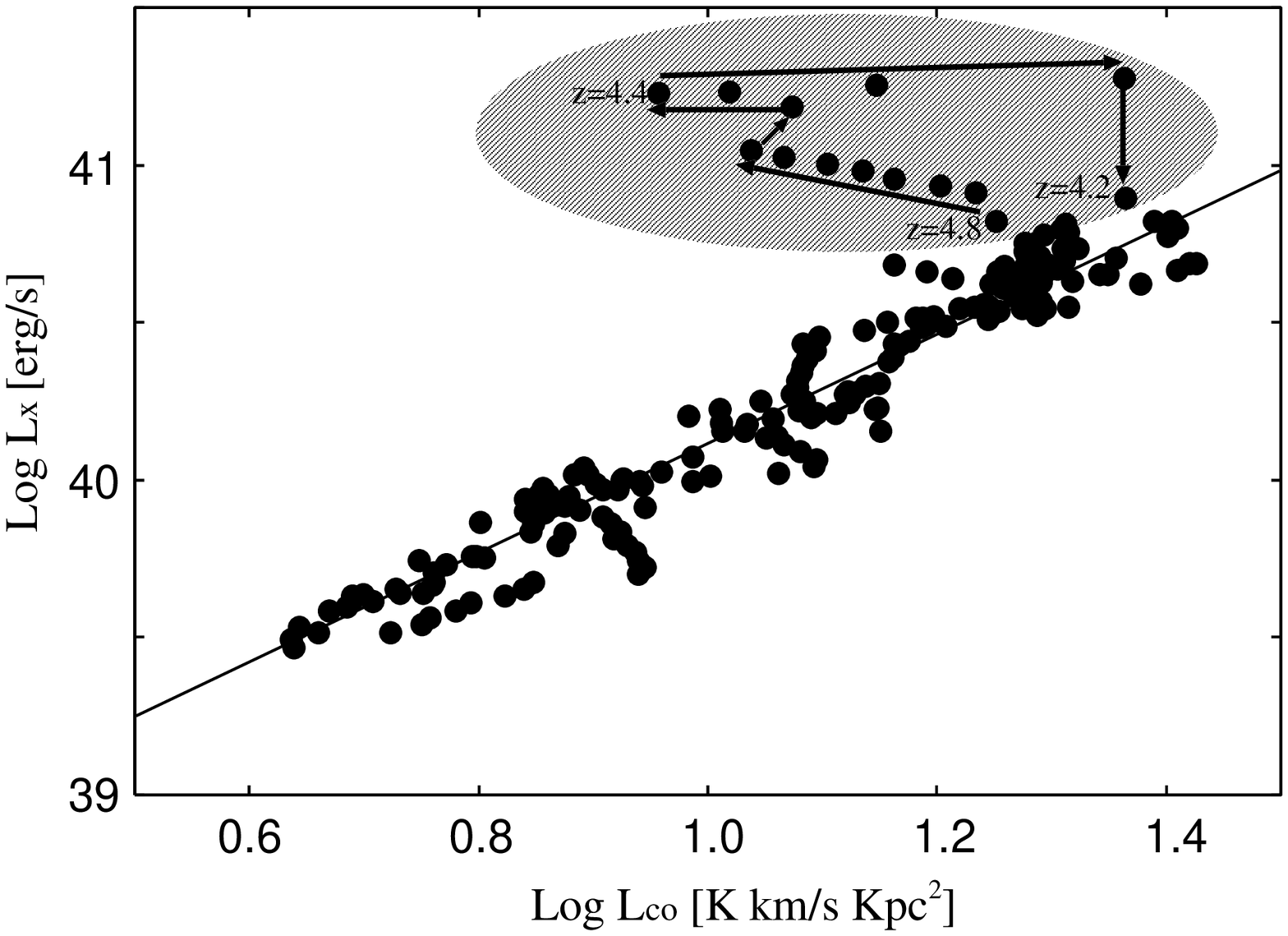}
\figcaption
{
The relation between the CO luminosity ($L_{\rm CO}$) and the X-ray 
luminosity ($L_{\rm X}$) during the evolution of the BH and the galaxy. The filled circles 
show time evolution in our model.
The X-ray luminosity is positively correlated with the CO luminosity. 
The thin line represents this correlation, i.e. $\log{L_{\rm X}}({\rm erg}/{\rm s})\sim 
38+2\log{L_{\rm CO}}({\rm K}\,{\rm km}\,{\rm s}^{-1}\, {\rm Kpc}^{2})$.
The shaded area corresponds to the BH growing phase ($4.2 < z < 4.8$). 
The arrows show time evolution in the BH growing phase.
}
%\label{Figure 7}
%\end{figure}

\subsection{$L_{\rm X}-L_{\rm CO}$ relation}
%[KW: THE FOLLOWING PAR. DO NOT MAKE SENSE FOR ME.]
In \S 3.3, we predicted that the X-ray luminosity of AGNs
is positively correlated with the CO luminosity of bulges from $z\sim 10$
to $z\sim 2$. This correlation (i.e. $L_{\rm X} \sim 10^{38} ({\rm ergs} \; {\rm s}^{-1}) \\(L_{\rm CO}/{\rm K}\, {\rm km}\, {\rm s}^{-1} \, {\rm kpc}^2)^2$),
 if we extrapolate it to AGNs with higher luminosity, is
consistent with the $L_{\rm X}-L_{\rm CO}$ relation
found in the low redshift Seyfert galaxies and quasars (Yamada 1994).
The observed $L_{\rm X}-L_{\rm CO}$ relation shows a scatter of 
about one order of magnitude.
Suppose all galaxies follow the same $L_{\rm
X}-L_{\rm CO}$ relation that we found here, 
we suspect that the large scatter in the observed scaling relation
was caused when the BHs were in their growing phases. 
%If this is the case, this trend is
%consistent with the results which the $L_{\rm X}-L_{\rm CO}$ relation
%found in the low redshift Seyfert galaxies and quasars (Yamada 1994).
Submillimeter observations by ALMA for luminous high-$z$ quasars
will be helpful to investigate the connection between the black hole 
mass and the host properties.
%Additional support for such a connection comes from the detection of 
%submillimeter CO emission in a number of extremely luminous high-$z$
%quasars and radio galaxies (Omont et al. 2003), although the current
%sensitivity of submillimeter telescope is insufficient to investigate
%this correlation for 
%less luminous AGNs at high-$z$. 

\section{Conclusions}
Combining a theoretical model of 
the mass accretion onto a galactic center due to the radiation drag 
with high-resolution $N$-body/SPH simulations ($2\times 10^{6}$
particles, one SPH particle has $10^{3}M_{\odot}$ and a softening length
of $\sim$ 50 pc), we demonstrate growth and formation of a
massive BH during hierarchical formation of a small spiral galaxy
(with a total mass of $10^{10}M_{\odot}$). 
We found that the average rate of the mass accretion due to the radiation drag 
is $\approx 10^{-5}M_{\odot}{\rm yr}^{-1}$.
% which is comparable to the
%Eddington mass accretion rate for a BH mass with $10^{4}M_{\odot}$, and
Finally, a small spiral galaxy can have an IMBH with $10^{4}-10^{5}M_{\odot}$ at
$z\sim 4$. 

Our model suggests that the growth of the massive BHs correlates not only 
with that of the galactic bulges, but also with that of the dark matter halos 
in the hierarchical formation of spiral galaxies.
The massive BHs coevolve with the dark matter halo from $z\sim 15$ to $z\sim 2$. 
This means that the change in the dark matter potential closely correlates
with the rate of the mass accretion onto a seed BH with the help of the radiation
drag. The final mass ratio of the BH-to-dark matter halo is $\approx
10^{-6}$ and the final BH-to-bulge mass ratio is about $5\times 10^{-5}$ 
in a small spiral galaxy, which is much smaller than the observed value ($\approx 10^{-3}$) in the large galaxies due to the opacity effect, although the stellar feedback would affect on the result. 
%The mass ratio of the BH-to-bulge is reduced by a factor of $\approx 10$, 
%compared with the observed value in the massive galaxies because of the
%opacity effects or the angular momentum problem in spiral galaxies.
Moreover, the time-lag between the BH growth and the growth of the bulge (halo) 
would cause the scatter of the observed scaling relation. 

In terms of relationship between the AGN activity and the properties 
of host galaxies, we found that even a small spiral galaxy could 
show the same level of the typical low luminosity AGNs with 
$L_{\rm AGN}/L_{\rm bulge}\approx 0.1$ 
at $z\approx 4$. 
Our model shows that the X-ray luminosity of 
the AGN is positively correlated with the CO luminosity (the gaseous
mass) of the bulge very well. Furthermore, our result predicts that the
BH-growing objects deviate from this scaling relation. 
%In order to establish this method, we will confirm if our prediction can
%be supported for NLS1s as future works.
%Moreover, 
By comparing our results with the properties of the LBGs, 
we predict that the LBGs could harbor massive BHs with $10^{6}-10^{7}M_{\odot}$.

\acknowledgments
The authors thank the anonymous referee for his/her fruitful comments and suggestions. NK acknowledges Italian MIUR and INAF financial support. Numerical computations were carried out on GRAPE clusters (MUV) and Fujitsu VPP 5000 at NAOJ (MUV Project ID g04a07, VPP Project ID rkw20a).The authors are supported by Grants-in-Aid for Scientific Research (no. 15684003 (KW) and no. 16204012 (TS)) of JSPS. 
%\newpage
%%%%%%%%%%%%%%%%%
% (5)References %
%%%%%%%%%%%%%%%%%

\end{document}